# Cross-Layer Design of Influence Maximization in Mobile Social Networks

Chih-Hang Wang[1], Po-Shun Huang[1], De-Nian Yang[2], and Wen-Tsuen Chen[1,2]
[1]Department of Computer Science, National Tsing Hua University, Hsin-Chu, Taiwan
[2]Institute of Information Science, Academia Sinica, Taipei, Taiwan
s100062591@m100.nthu.edu.tw, s102062626@m102.nthu.edu.tw, dnyang@iis.sinica.edu.tw, chenwt@iis.sinica.edu.tw

*Abstract*—Most prior algorithms for influence maximization focused are designed for Online Social Networks (OSNs) and require centralized computation. Directly deploying the above algorithms in distributed Mobile Social Networks (MSNs) will overwhelm the networks due to an enormous number of messages required for seed selection. In this paper, therefore, we design a new cross-layer strategy to jointly examine MSN and mobile ad hoc networks (MANETs) to facilitate efficient seed selection, by extracting a subset of nodes as agents to represent nearby friends during the distributed computation. Specifically, we formulate a new optimization problem, named Agent Selection Problem (ASP), to minimize the message overhead transmitted in MANET. We prove that ASP is NP-Hard and design an effectively distributed algorithm. Simulation results in real and synthetic datasets manifest that the message overhead can be significantly reduced compared with the existing approaches.

## I. Introduction

Previous research has indicated that people tend to accept information shared by their friends more than advertisements presented in news or on TV [1]. With the growing popularity of social networks such as Facebook and Twitter, social networks currently play an important role in the spread of information or influence in the form of "word-of-mouth" [2], which has drawn increasing attention recently [3][4][5]. To facilitate viral marketing, a company usually offers the free samples of a product to a "small" group of seed users according to the budget. If these users like the product, they are inclined to recommend their friends to adopt it, and their friends may influence their friends' friends to give it a try and so on. Through this powerful chain effect of influence, many individuals will eventually adopt this product. Since the company wishes that the initial seeds will ultimately influence more people in the network, how to select these initial people to try the samples is crucial.

Formally, the problem is defined as *influence maximization* [6][7][8][9]. The goal is to find $K$ seed nodes for maximizing the *spread*, which is the number of *active* nodes (e.g., the people that like or purchase the products). Kempe et al. [6] first proved that the problem is NP-hard and designed a $1 - 1/e$ approximation algorithm, which was further improved in terms of running time in [10][11]. Qin et al. [12] studied the mobile advertising in vehicular networks. Liu et al. [13] grouped users and found the users with the highest influence on each topic. However, all previous influence maximization algorithms [3]-[13] require centralized computation and are not designed for distributed environments, such as mobile social networks (MSNs).

Currently, mobile devices can communicate with each other through Bluetooth and WiFi-Direct to construct Mobile Ad Hoc Networks (MANETs). Many mobile social applications have been developed (e.g., Micro-blog [14] and SocialbleSense [15]) to build MSNs on MANETs, and related security and privacy issues have been studied [16], [17]. Also, Ning et al. [18] proposed a Self-Interest-Driven incentive scheme to encourage cooperation in data dissemination on MSNs. For energy efficiency, Hu et al. [19] proposed an energy-aware user contact detection to detect user movement in MSNs. Lu et al. [20] formulated a new minimization problem for information diffusion as an asymmetric *k*-center problem. Nevertheless, distributed influence maximization for MSNs on MANETs has not been explored.

New challenges arise for efficiently influence maximization in MSNs. First, most existing algorithms [6] [10] [12] [20] require centralized computation. Nevertheless, due to the privacy setting in most mobile social APPs, [21][22], no user is able to acquire the topology of the whole MSN, and each user can only know the identities of her friends or friends of friends. In other words, it is not possible to deploy a centralized algorithm in a single node for finding the seeds in MSN. Second, for cross-layer optimization, a nearby friend in the MSN may be far away in the MANET. Therefore, it is envisaged that an efficient cross-layer strategy to minimize message overhead by considering both networks is desired.

In this paper, therefore, we first study the cross-layer design of distributed influence maximization for MSNs on MANETs. To efficiently find the seed nodes in MSN, it is crucial to avoid overwhelming MANET with information exchange during the distributed computation. To achieve the goal, our idea is to select a set of *agents* in MANET, where each agent is a representative of a set of her friends in MSN to participate the distributed computation of the seed selection. Messages are exchanged only between the agents in MANET, instead of between any two friends in traditional algorithms. Therefore, it is envisaged that a great number of information messages can be reduced.

More specifically, we formulate a new optimization problem, named Agent Selection Problem (ASP), to select an agent for each user in the MSN to minimize the total message overhead transmitted in MANET. We prove that ASP is NP-Hard and then design a distributed algorithm, called Agent Selection for Minimizing Transmission Cost (ASMTC), for ASP. ASMTC iteratively extracts the agent reducing the most message overhead in the MSN, whereas the distance (i.e., the number of hops) between two agents in MANET is also examined. Simulation results in real and synthetic datasets [23] show that our algorithm can significantly reduce message overhead to support distributed seed selection in MSN.

The rest of paper is organized as follows. Section II introduces the information diffusion model, the distributed seed selection, and the concept of message overhead. We present the problem formulation and prove the hardness result. In section III, the distributed algorithm is proposed. Simulation results are presented in section IV. Finally, section V concludes the paper.

## II. SYSTEM MODEL AND PROBLEM STATEMENT

In this section, we first introduce the influence diffusion model and then describe the message overhead for deploying the most popular approximation algorithms [4][6][7] in MSN. Next, we formulate the Agent Selection Problem (ASP) and present the hardness result.

### A. Influence diffusion model

Independent Cascade (IC) [6][11][24] is a popular stochastic model for influence diffusion in social networks. Each node is either *active* or *inactive* in IC. The state can only be switched from *inactive* to *active* according to the following procedures. 1) The influence propagation starts from the seeds in round $r = 0$ initially. Let $S_0$ be the set of *active* nodes with only the seeds. 2) In the following rounds, any node $u \in S_{r-1}$ has one chance to influence each of its *inactive* neighbors by an independent probability $p_v$, whereas $v \in N(u)$, and $N(u)$ is the set of neighbors of $u$. If neighbor $v$ is activated by $u$, the state of $v$ is switched from *inactive* to *active*, and $v$ is added to $S_r$. Note that $S_r$ is set as $\emptyset$, instead $S_{r-1}$ in the beginning of each iteration. 3) The process repeats until $S_r$ is empty. In other words, no node can be activated in the future. Finally, the *spread* is the set of *active* nodes.

Fig. 1(a) presents an illustrative example. When node 4 is the seed, $S_0 = \{4\}$, node 4 is initially set to be *active*. Next, node 4 influences its neighbors $N(4) = \{2,3,6,7\}$ by an independent probability $p_v$. Assume that nodes 2, 3 and 6 are successfully influenced by node 4 and thereby switched to *active* and added to $S_1$. The neighbor of nodes 2, 3, and 6 are nodes 1, 4, and 6, nodes 1, 4, and 8, and nodes 1, 2, and 4, respectively, where nodes 2, 3, 4, and 6 are already active. Therefore, nodes 2 and 6 try to influence node 1, whereas node 3 tries to influence nodes 1 and 8 in this round. Assume that neither node 1 nor node 8 are successfully influenced, and therefore, no node is attached to $S_2$. Finally, the algorithm stops since $S_2$ is empty, and the influence spread includes nodes 2, 3, 4, and 6.

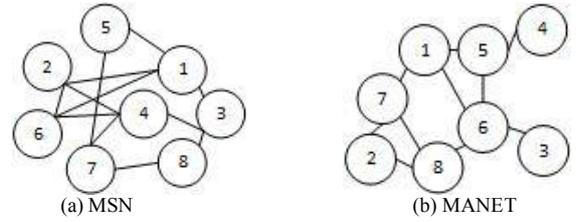

Figure 1. Example of MSN and MANET

### B. Distributed Seed Selection in MSN

The IC model describes the influence diffusion after the seeds are selected. For seed selection in MSN, a distributed algorithm is required because no node has the topology of MSN. A promising way to deploy the current popular approximation algorithm [6] in MSN is described as follows. Initially, each node $v$ assumes that the initial seed set $S_1 = \{v\}$ and starts to influence her friends with the IC model. When any node becomes active, it returns its state to $v$, and $v$ acquires its influence spread and broadcasts the spread size to every other node. Assume that $u$ is the node leading to the largest spread, and therefore $S_1 = \{u\}$ in the end of the first iteration. Then, each node $w$ assumes that $S_2 = \{u, w\}$ and starts to influence her friends similarly, whereas $S_2$ is included in the message to neighbors. For node $u$, it is involved in set $S_2 = \{u, w\}$ for every $w$, and each set tries to influence a friend once (because the spread of each different $S_2$ needs to be calculated individually). Every active node for each set $S_2 = \{u, w\}$ returns its state to $w$, and every node $w$ acquires its influence spread of $\{u, w\}$ and broadcasts the spread size to every other node. Assume that $v$ is the node with the largest spread, and therefore, $S_2 = \{u, v\}$ in the second iteration. The above process repeats until $S_K$ is acquired, where $K$ is the budget parameter.

Apparently, the above distributed seed selection involves many message exchanges due to the following reasons. 1) In every iteration $i$, every node $w$ not in $S_{i-1}$ needs to find the spread with the seed set $S_i = S_{i-1} \cup \{w\}$. In other words, message exchanges are necessary to be involved for different $w$ to find the corresponding spread size. 2) Most importantly, the spread size of each seed set needs to be derived multiple times in order to find the average size [6] in this probabilistic IC model since finding the exact spread size analytically is #P-Complete [6] (i.e., not feasible).

### C. Message Overhead in MANET

In MSN, the influence message from a node to her friend may traverse multiple hops in MANET, and the message overhead [25] thereby is defined as the total number of hops for all messages. For example, Fig. 1(b) considers the case with two messages delivered from node 4 to 6 and node 3 to 4, respectively. The number of hops is 2 from node 4 to node 6 via node 5, whereas the number of hops is 3 from node 3 to 4 via nodes 6 and 5. Therefore, the total message overhead is $2 + 3 = 5$. It is envisaged that directly applying the above approximation algorithm in a distributed environment will incur a large message overhead.

In this paper, therefore, we first show that the message overhead can be significantly reduced by selecting an agent for each node in MANET. Since an individual's friends in MSN may be distant to him in MANET, the influence message needs to traverse many hops to influence the friends. Therefore, it is expected when an agent who is closer to the individual's other friends acts as a representative[1], the message overhead can be significantly reduced. Consider the illustrative example in Fig. 1, where the spread is $S_0 \cup S_1 \cup S_2 = \{2,3,4,6\}$ with node 4 as the initial seed. In the traditional influence diffusion, the total message overhead is 9 with 4 messages to influence node 2, 3 messages to influences node 3, and 2 messages are involved in influencing node 6. Notice that the message is transmitted only when the neighbor is successfully influenced according to the successful probability. Since nodes 1 and 8 are not successfully influenced, there is no message sent to them.

In contrast, consider the case when node 4 is the agent for nodes 2, 4, and 6, and node 1 is the agent for nodes 1, 3, and 5. Since node 4 is the agent for nodes 2 and 6, the message overhead of node 4 needed for influencing nodes 2 and 6 is 0 (i.e., it can be locally computed and decided according to successful probability). In other words, node 4 acts on behalf of nodes 2 and 6 for transmitting and receiving messages. On the other hand, the message overhead is 2 for node 4 to influence node 3 because node 3's agent is node 1, and node 4 transmits the influence message to node 1 in two hops, instead of transmitting it to a more distant node 3. The message overhead for influencing node 1 and 8 is 0 since they are not successfully influenced. Therefore, the total message overhead is 2, and it is much smaller than the conventional approach (i.e., 9).

Note that the agent of a node $u$ (called $agent(u)$) needs to send an influence message to the agent of $v$ (called $agent(v)$) if 1) $agent(u)$ and $agent(v)$ are different nodes, and 2) $u$ indeed activates $v$ according to the successful probability $p_{u,v}$. In contrast, if $u$ fails to activate $v$ according to the successful probability $p_{u,v}$, it is not necessary for $agent(u)$ to send a message to $agent(v)$. In other words, in the derivation of the spread, whether an influence trial is successful can be decided by $agent(u)$, instead of $agent(v)$, to reduce the message overhead.

*D. Problem Statement*

Based on the above observation that jointly examines MSN and MANET, we formulate a new problem, named Agent Selection Problem, as follows. Let $G^S = (V^S, E^S)$ represent the MSN with each vertex $v$ in $V^S$ as an individual and each directed edge $e$ in $E^S$ as the friendship connection. Each edge from $u$ to $v$ is associated with a successful probability $p_{u,v}$, and $p_{u,v}$ can be different from $p_{v,u}$. On the other hand, let $G^A = $

---

[1] Note that the agent is involved only in the derivation of the spread in the algorithm, which is exploited to quantify and evaluate the effectiveness of the seeds to solve the spread maximization problem (as a prediction). When the person-to-person word-of-mouth viral marketing starts, each person will make her own decision based on the influence from friends, and no agent will be involved as a representative of friends for decision making.

$(V^A, E^A)$ denote the MANET with each vertex $v$ in $V^A$ as an individual. An edge $(u, v)$ exists in $E^A$ if $u$ and $v$ are within the transmission range of each other. In other words, each individual $v$ can only directly communicate with her neighbor set $N^A(v)$ in MANET, whereas multi-hop relay is necessary to be involved if her friends in the neighbor set $N^S(v)$ of MSN are not in $N^A(v)$. Specifically, ASP is formulated as follows:

**Problem:** The Agent Selection Problem
**Instance:** An MSN $G^S = (V^S, E^S)$ with successful probability $p_{u,v}$ associated with each edge $(u,v)$ in $E^S$, a MANET $G^A = (V^A, E^A)$.
**Task:** To find the agent for each individual such that the total message overhead is minimized.

In the following, we prove that ASP is NP-Hard with the reduction from Minimum-Weight Dominating Set Problem (MWDSP).

**Theorem 1.** ASP *is NP-hard*

**Proof.** Let graph $G = (V, E)$ be an instance of MWDSP, where each vertex $v \in V$ is associated with a weight $w_v$. The problem is to find a subset $D$ of $V$ such that every vertex $v \notin D$ is adjacent to at least one vertex $u \in D$ and the total weight $\sum_{u \in D} w_u$ is minimized. For graph $G$, we construct an instance of ASP by defining $G^S = (V^S, E^S) = G$ as the topology of nodes in the MSN where each node represents an individual. Let $V^S$ be the set of nodes and $E^S$ be the set of connections of nodes where the connection $(x, y) \in E^S$ exists if $x$ and $y$ are friends in the MSN and $x, y \in V^S$. Let $c_x$ be the cost of node $x$ when $x$ acts as an agent for her friends to influence other friends of $x$ in the MSN. The calculation of $c_x$ is based on the number of hops needed for two nodes to communicate with each other in MANET. Since $c_x$ may vary due to the selection of agents from the other individuals, we set the value of cost such that no node has selected any agent as $c_x$, that is, each node influences her friends by herself.

For MWDSP, if there exists a subset $D$ of $V$ such that every vertex $v \notin D$ is adjacent to at least one vertex $u \in D$, we can find the corresponding node subset $A$ of $V^S$ such that for each node $x \notin A$, at least one of her friends is selected as an agent for $A$ where $A$ is the set of selected agents. Conversely, if there exists a node subset $A$ of $V^S$ such that for each node $x \notin A$, at least one of her friends is selected as an agent for $A$, we can find out the corresponding subset $D$ of $V$ such that every vertex $v \notin D$ has an edge connected to at least one vertex $u \in D$. This is because each node $x \in V^S$ with cost $c_x$ corresponds to a vertex $v \in V$ with weight $w_v$ and each connection $(x, y) \in E^S$ corresponds to an edge $(u, v) \in E$. Notice that in ASP, each node can only select at most one of her friends as an agent who acts on behalf of her. Therefore, after selecting a node subset $A$ of $V^S$, we select, for each node $x \notin A$, one agent for $A$ who induces the least cost on behalf of $x$. Hence, the hardness of ASP is as hard as WSCP; the theorem follows.

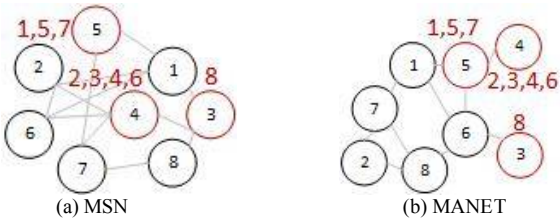

(a) MSN  (b) MANET

Figure 2: The final result of agent selection

### III. ALGORITHM DESIGN

In this section, we first introduce the concept of algorithm ASMTC in Section III.A and then explain the details in Section III.B. Time complexity is analyzed in Section III.C.

*A. Algorithm Concept*

To solve ASP, we propose Agent Selection for Minimizing Transmission Cost (ASMTC) to select an agent for each node to minimize the total message overhead in MANET. In the following, the nodes that have not selected the agents are *unrepresented* nodes, whereas the others are *represented* nodes. ASMTC first determines the candidate agents for each node. Then, the algorithm iteratively selects the node that can reduce the most message overhead as the representatives for its *unrepresented* friends. Afterward, ASMTC further reduces the message overhead by adjusting the agent of each node if the new agent can represent the node with a smaller message overhead.

More specifically, ASMTC consists of two phases: 1) Distributed Agent Selection (DAS), and 2) Message Overhead Reduction (MOR). In the following, we first describe several design objectives behind ASMTC. First, it is crucial to select the agents for representing neighboring nodes, by jointly considering the topologies of MSN and MANET. The reason is that a node's friends may not be around her in MANET and therefore, the direct communication between them may traverse many hops[2]. Moreover, larger overhead reduction is inclined to be achieved if an agent can represent more nearby friends, because the influence diffusion of those friends can be computed by the agent directly, and no message overhead will be incurred in this case. To acquire a good solution, we propose an ordering scheme to iteratively select the agents and design a local agent adjustment method to further reduce the computational overhead and message overhead for MANET[3].

*B. Algorithm Details*

In the following, we describe ASMTC in detail.

*1) Distributed Agent Selection (DAS)*

---

[2] The mobility of nodes can be managed as that in [26] since it traces nodes by constructing a virtual backbone of networks in a distributed manner.
[3] To avoid increasing computational overhead when networks become larger, parameters $\alpha$ and $\beta$ are included to limit the number of hops of message transmission and to limit the times for adjusting the agent of each node, respectively.

In this phase, since ASMTC aims to minimize message overhead, ASMTC iteratively selects the node that can reduce the largest message overhead to act as representatives for its *unrepresented* friends or the node itself. Notice that node $v$ can also act as its agent, and the messages are transmitted from agents to agents. The default agent for each node $v$ is $v$ itself. First, each *unrepresented* node $v$ estimates the reduced message overhead of $v$ (called $RMO(v)$) under the circumstances that $v$ is $agent(v)$ and represents each *unrepresented* node $u \in N^S(v)$. Then, node $v$ asks each $u$ to calculate $RMO(u)$ under the circumstances that node $v$ is $agent(u)$, and the result is returned to $v$. Afterward, node $v$ calculates $RMO(v) + \sum RMO(u)$ and then broadcasts the result to every other node. Notice that the reduced message overhead is the difference between the message overhead before and after selecting an agent, and therefore the value may be negative (i.e., the message overhead increases after selecting the agent). Assume that $w$ is the node reducing the largest message overhead if it is the agent and therefore $A = \{w\}$ in the end of the first iteration, where $A$ is the agent set. Next, for each *unrepresented* node $v$, $v$ asks each *unrepresented* $u \in N^S(v)$ to calculate $RMO(u)$ under the circumstances that $v$ is $agent(u)$, and the result is returned to $v$. Notice that even though a node is *represented,* it can still represent other nodes. Therefore, we consider one more case after the first iteration. For each *represented* node $v$, if there is still *unrepresented* node $u \in N^S(v)$, $v$ asks each *unrepresented* $u \in N^S(v)$ to calculate the reduced message overhead under the circumstances that $v$ is $agent(u)$, and then $u$ returns the result to $v$. Afterward, node $v$ calculates $\sum RMO(u)$ and then broadcasts the result to every other node. Assume that $x$ is the node reducing the largest message overhead if it is the agent, and therefore $x$ is involved in $A = \{x, w\}$ in the end of the second iteration. After determining an agent, each node $v$ notifies each node $u \in N^S(v)$ of $agent(v)$'s identity, and $u$ then relays this information to $agent(u)$. Also, each node $v$ notifies the identities of $N^S(v)$ and their corresponding agents to $agent(v)$. DAS stops when every node has an agent.

Fig. 2 presents an illustrative example. To simplify the presentation, we only explain the detailed behavior of a few nodes, and the process of the other nodes is similar and is omitted due to the space constraint. At the beginning, the total message overhead for all nodes is 50. For node 4, the message overhead is $4 + 3 + 2 + 3 = 12$ to influence nodes {2,3,6,7}. If node 4 becomes the agent of nodes {2,3,4,6,7}, the reduced message overhead is 12 since node 4 represents all its friends. Then, nodes {2,3,6,7} calculate the reduced message overhead individually under the circumstances that node 4 is the agent of them. The reduced message overhead of nodes {2,3,6,7} is {6,2,3,2}, and therefore, the reduced message overhead for selecting node 4 as the agent of all its friends is $12 + 6 + 2 + 3 + 2 = 25$. Afterward, node 4 broadcasts the reduced message overhead to every other node. In this example, node 4 is just the node with the largest reduced message overhead in the first iteration of DAS. The final result in DAS is that node 4 represents nodes {2,3,4,6,7}, node 5 represents nodes {1,5}, and

**Algorithm 1.** ASMTC

1: **Initialize**:
2: **for** each $v \in V^S$ **do**
3:    $RMO(v) = 0$
4:    $agent(v) = v$
5:    $cost_v = 0$
6:    $v$ is *unrepresented*
7: **end for**

**Distributed Agent Selection (DAS):**
8: **repeat**
9:   **for** each $v \in V^S$ **do**
10:     **if** $v$ is *unrepresented* **then**
11:       $v$ calculates $RMO(v)$ with $agent(v) = v$
12:     **end if**
13:     **for** each $u \in N^S(v)$ && $u$ is *unrepresented* **do**
14:       $u$ calculates $RMO(u)$ with $agent(u) = v$
15:       $u$ returns the result to $v$
16:     **end for**
17:     $cost_v = RMO(v) + \sum RMO(u)$
18:     $v$ broadcasts $cost_v$ to every other node
19:     $v$ compares $cost_v$ with every $cost_u$ from each $u \in V^S$
20:   **end for**
21:   the node $w$ with the largest $cost_w$ is elected as the agent
22:   **if** $w$ is *unrepresented* **then**
23:     $agent(w) = w$
24:     $w$ is *represented*
25:   **end if**
26:   **for** each $u \in N^S(w)$ && $u$ is *unrepresented* **do**
27:     $agent(u) = w$
28:     $u$ is *represented*
29:   **end for**
30:   every node broadcasts its agent to its friends
31: **until** every $v \in V^S$ is *represented*

**Message Overhead Reduction (MOR):**
32: **repeat**
33: *trying*:
34:   **for** each $v \in V^S$ **do**
35:     **for** each $u \in CA(v)$ && $u \neq agent(v)$ **do**
36:       $v$ calculates $RMO(v)$ with $agent(v) = u$
37:       **if** $RMO(v) > 0$ **then**
38:         $agent(v) = u$
39:       **end if**
40:       $v$ broadcasts $agent(v)$ to its friends
41:     **end for**
42:   **end for**
43: *checking*:
44:   **for** each $v \in V^S$ **do**
45:     **if** $RMO(v) < 0$ **then**
46:       $v$ broadcasts $RMO(v)$ to every other node
47:     **end if**
48:   **end for**
49:   every node checks the total message overhead.
50:   **if** the total message overhead is reduced **then**
51:     every node adjusts the agent determined in *trying*
52:     **continue**
53:   **else then**
54:     *backward tracking*
55:   **end if**
56: *backward tracking:*
57:   **repeat**
58:     the node $w$ with the smallest $RMO(w)$ asks $v \in N^S(w)$ that increases the largest message overhead of $w$ to maintain the agent.
59:     *checking*
60:   **until** the total message overhead is reduced
61:   every node selects the agent determined in *checking* and *backward tracking*.
62: **until** the total message overhead cannot be reduced

---

node 3 represents node 8. The total message overhead for influencing friends reduced from 50 to 20.

*2) Message Overhead Reduction (MOR)*

In the MOR phase, we further reduce the overall message overhead by locally adjusting the agent of each node. MOR has three stages, *trying*, *checking*, and *backward tracking*. Notice that the following calculation of each node is done by the agent. In the *trying* stage, each node $v$ tries each candidate agent $u \in N^S(v)$, except the current agent (i.e., the agent selected in DAS), to be $agent(v)$. Note that the candidate agent of $v$ can only be $v$ itself or $v$'s friends. Then, each $v$ calculates $RMO(v)$ under the circumstances that $u$ is $agent(v)$ and chooses the node that can reduce the largest message overhead as the agent and broadcasts the selection result to its friends. Again, notice that the reduced message overhead may be negative, and therefore, node $v$ will not change its agent if the reduction is negative. Afterward, MOR enters *checking* stage. In the *checking* stage, each node $u \in N^S(v)$ estimates $RMO(u)$ (here is the difference between the message overhead before and after $v$ adjusts the agent) with the agent determined in the *trying* stage. If $RMO(u)$ is negative, node $u$ broadcasts $RMO(u)$ to every other node, and every other node also broadcasts its reduced message overhead. Then, the total reduced message overhead over the whole network is acquired by every node, and each node $v$ selects the agent determined in the *trying* stage if the message overhead over the whole network is still reduced.

On the other hand, if the total message overhead increases, MOR enters *backward tracking* stage. In this stage, the node $u$ with the largest increased message overhead first requests the friend $v \in N^S(u)$ that increases the largest message overhead of $u$ to maintain the original agent, and then every node again estimates the reduced message overhead as described in the *checking* stage. The above process repeats until the total message overhead is reduced, and then each node $v$ selects the agent determined in the *checking* and *backward tracking* stages. With the help of *backward tracking* stage, the total message overhead can be further reduced in each iteration. MOR stops when the total message overhead cannot be further reduced.

TABLE I. PARAMETER SETTINGS

| Parameter | Value | Means |
|---|---|---|
| N | 15233,37154 | number of nodes n |
| $\alpha$ | 10,45 | average degree |
| $\alpha_{max}$ | 15,50 | maximum degree |
| $\xi_1$ | 2 | minus exponent for the degree |
| $\xi_2$ | 1 | minus exponent for the community size distribution |
| $\mu_t$ | 0.1 | mixing parameter for the topology |
| $\mu_w$ | 0.1 | mixing parameter for the weight |

We consider node 7 in Fig. 2 as an example. Notice that the agent selection result in the DAS is used. Node 7 first tries its candidate agents {5,7,8} and calculates the reduced message overhead of node 7, and the results are {1, −4, −3} in the *trying* stage. Then, node 7 chooses node 5 to be its agent because it can reduce the largest message overhead, and MOR enters *checking* stage after all nodes try their agents. In the *checking* stage, node 4 observes that its message overhead increases 1 after node 7 adjusts its agent to node 5. However, the total message overhead still decreases from 20 to 18, and therefore node 7 adjusts its agent to node 5. Finally, ASMTC stops since no message overhead of nodes can be further reduced.

Fig. 2(b) illustrates the final agent for each node. The nodes selected as the agents are circled in red, whereas the numbers next to them are the nodes they represent. It is worth noted that the agents are close to each other in the MANET and each agent represents several her friends, and therefore, selecting them as the agents significantly reduces the message overhead and much influence diffusion of their friends can be computed by the agents directly without any message transmission.

*C. Time complexity*

In this section, we analyze the time complexity of ASMTC for each node $v$. In the DAS, each node $v$ calculates the reduced message overhead of $v$ under the circumstances that $v$ represents $v$ and its friends $N^S(v)$ and then compares it to the other nodes. The worst case is that all the nodes are connected, and in this case, each node considers the other $|V^s| - 1$ nodes to compute the reduced message overhead and compares its reduced message overhead with them. It takes at most $O(|V^s|^2)$ times in the DAS. In the MOR, each node tries to adjust its agent in the *trying* stage, and the worst case is that the node tries $|V^s| - 1$ nodes. Therefore, it takes $O(|V^s|)$ times to try each node in the *trying* stage. In the *checking* stage, the time complexity is $O(|V^s|)$ since each node examines the reduced message overhead from other nodes and checks whether the value is negative or not. In the *backward tracking* stage, the worst case is that all the nodes go back to maintain the agents, and it takes at most $O(|V^s|)$ times. As a result, the time complexity of MOR is $O(|V^s|)$, and the overall time complexity of ASMTC is $O(|V^s|^2 + |V^s|)$.

IV. SIMULATION

In this section, we compare the seed selection algorithm with and without ASMTC under two different real datasets.

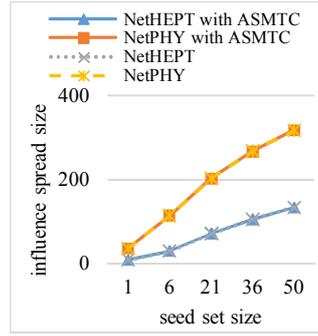
Figure 3: Influence spread

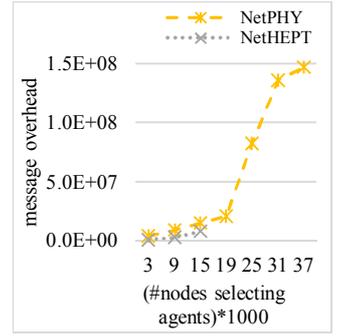
Figure 4: Overhead of ASMTC

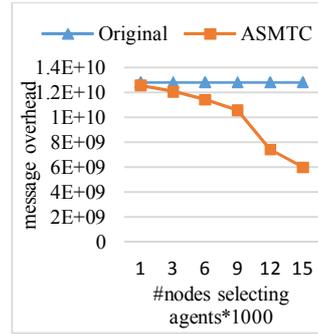
(a) OSN NetHEPT

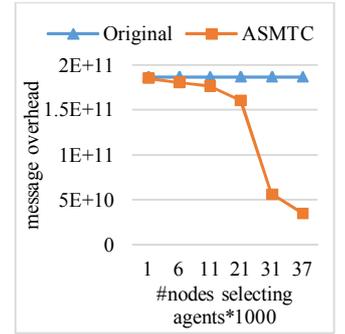
(b) OSN NetPHY

Figure 5: Varying number of nodes selecting agents in NetHEPT and NetPHY

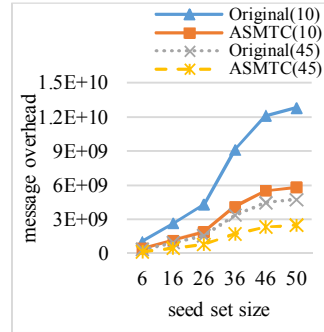
(a) NetHEPT

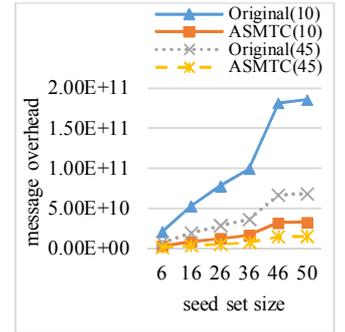
(b) NetPHY

Figure 6: Message overhead in NetHEPT and NetPHY

*A. Simulation Setups*

We evaluate ASMTC with two real social networks from arXiv5[4], which were also used in the simulations in [6] and [11]. The first network (NetHEPT) includes 15,233 nodes and 58,891 edges. The second one (NetPHY) contains 37,154 nodes and 231,584 edges. For MANET, we exploit the synthetic networks [27] with the detailed setting the same as [20] and listed in Table I, where the number of nodes is the same as MSN. Each node in the MSN is randomly mapped to a node in the MANET. To find the seeds, we adopted the CELFGreedy [10] algorithm, which can effectively reduce the running time of the popular approximation algorithm [6] by 700 times.

---
[4] http://www.arXiv.org

*B. Simulation Results*

Fig. 3 first presents the results of CELFGreedy [10] with and without ASMTC. The solutions from CELFGreedy with and without ASMTC are the same because the role of ASMTC is only to facilitate distributed computation of CELFGreedy with the selected agents to reduce the message overhead in MANET. Fig. 4 evaluates the message overhead required by ASMTC to select the agents. The result indicates that the additional overhead from ASMTC is relatively small compared with the total overhead. In Fig. 4, the amount of message overhead caused by ASMTC is at most $1.5 \times 10^8$, whereas the message overhead is reduced by more than $10^{11}$ as shown in Fig. 5(b). It is because ASMTC is only required to perform one time, where an enormous number of spreads is required to be generated and compared for different combinations of seeds in CELFGreedy. Fig. 5 displays the results with different numbers of nodes allowed to exploit agents, where the seed number is 50. ASMTC runs until the target number of nodes selecting agents is reached, while the remaining nodes will not select agents. Note that the original CELFGreedy algorithm does not employ any agent, and thus the corresponding message overhead remains the same. In contrast, Fig. 5 manifests that a small number of agents is able to effectively reduce the overhead.

Fig. 6 evaluates the message overhead of ASMTC with different average degrees (10 and 45) in MANET. The results manifest that more overhead is generated when the spread size grows because more candidate seeds are required to be investigated, and more nodes are involved in the calculation of a spread. With a larger degree in MANET, each node has more chances to find a shorter path to a destination, and the message overhead thereby decreases. Moreover, Fig 6 manifests that the overhead can be effectively reduced by 82% with the average degree as 15 and 77% with the average degree as 45, because ASMTC properly selects a smaller number of agents as representatives of friends to facilitate distributed seed selection.

## V. CONCLUSION

To the best of our knowledge, this paper is the first attempt to explore the reduction of message overhead for MSN influence maximization in MANET. We formulate a new optimization problem ASP and design an effective distributed algorithm ASMTC with the idea to exploit agents as the representative for distributed seed selection. We jointly consider the topologies of MSN and MANET when selecting agents and let an agent represents more nearby friends. To acquire a good solution, we propose an ordering scheme to iteratively select the agents and design a local agent adjustment method to further reduce the computational overhead and message overhead for MANET. The simulation results manifest that the message overhead can be effectively reduced by 52% - 82% with real datasets with a small number of agents.

## REFERENCES

[1] J. Nail, "The consumer advertising backlash," *Forrester Research and Intelliseek Market Research Report*, 2004.

[2] I. R. Misner, "The World's best known marketing secret: Building your business with word-of-mouth marketing," *Bard Press*, 1999.

[3] D. T. Nguyen, S. Das, and M. T. Thai, "Influence maximization in multiple online social networks," *Proceedings of IEEE Global Communications Conference (GLOBECOM)*, 2013.

[4] X. Fan, and V. OK Li, "Combining intensification and diversification to maximize the propagation of social influence," *Proceedings of IEEE Communications(ICC)*, 2014.

[5] V. Kumar, A.N. Sharma, N. Taneja, and H.Sharma "Optimal strategy for determining pricing policy of a product under the influence of reworking during production," *Proceedings of IEEE Conference on Computer Communications (IEEE INFOCOM)*, 2014.

[6] D. Kempe, J. Kleinberg, and E. Tardos, "Maximizing the spread of influence through a social network," *Proceedings of the ninth ACM SIGKDD international conference on Knowledge discovery and data mining (ACM KDD)*, 2003.

[7] F. Hao, M. Chen, C. Zhu, and M. Guizani, "Discovering influential users in micro-blog marketing with influence maximization mechanism," *Proceedings of IEEE Global Communications Conference (GLOBECOM)*, 2012.

[8] S. Li, Y. Zhu, D. Li, D. Kim, H. Ma, and H. Huang, "Influence maximization in social networks with user attitude modification," *Proceedings of IEEE Communications(ICC)*, 2014.

[9] J. Ok, J. Shin, and Y. Yi, "On the progressive spread over strategic diffusion: Asymptotic and computation," *Proceedings of IEEE Conference on Computer Communications (IEEE INFOCOM)*, 2015.

[10] J. Leskovec, A. Krause, C. Guestrin, C. Faloutsos, J. VanBriesen, and N. Glance, "Cost-effective outbreak detection in networks," *Proceedings of the 13th ACM SIGKDD international conference on Knowledge discovery and data mining (ACM KDD)*, 2007.

[11] W. Chen, Y. Wang, and S. Yang, "Efficient influence maximization in social networks," *Proceedings of the 15th ACM SIGKDD international conference on Knowledge discovery and data mining (ACM KDD)*, 2009.

[12] J. Qin, H. Zhu, Y. Zhu, L. Lu, G. Xue, and M. Li, "POST: exploiting dynamic sociality for mobile advertising in vehicular networks," *Proceedings of IEEE Conference on Computer Communications (IEEE INFOCOM)*, 2014.

[13] W. Liu, Z.-H. Deng, L. Cao, X. Xu, H. Liu and X. Gong, "Mining top K spread sources for a specific topic and a given node," *IEEE Transactions on Cybernetics*, vol. 45, no. 11, pp. 2472-2483, Nov. 2015.

[14] S. Gaonkar, J. Li, R. R. Choudhury, L. Cox, and A. Schmidt, "Micro-blog: sharing and querying content through mobile phones and social participation," *Proceedings of the 6th international conference on Mobile systems, applications, and services (ACM MobiSys)*, 2008.

[15] K. K. Rachuri, C. Mascolo, M. Musolesi, and P. J. Rentfrow, "SociableSense: exploring the trade-offs of adaptive sampling and computation offloading for social sensing," *Proceedings of the 17th annual international conference on Mobile computing and networking(ACM Mobicom)*, 2011.

[16] W. Dong, V. Dave, L. Qiu, and Y. Zhang, "Secure friend discovery in mobile social networks," *Proceedings of IEEE Conference on Computer Communications (IEEE INFOCOM)*, 2011.

[17] M. Li, N. Cao, S. Yu, and W. Lou, "FindU: privacy-preserving personal profile matching in mobile social networks," *Proceedings of IEEE Conference on Computer Communications (IEEE INFOCOM)*, 2011.

[18] T. Ning, Z. Yang, H. Wu, and Z. Han, "Self-interest-driven incentives for ad dissemination in autonomous mobile social networks," *Proceedings of IEEE Conference on Computer Communications (IEEE INFOCOM)*, 2013.

[19] W. Hu, G. Cao, S. V. Krishanamurthy, and P. Mohapatra, "Mobility-assisted energy-aware user contact detection in mobile social networks,"

*Proceedings of IEEE Conference on Distributed Computing Systems (ICDCS)*, 2013.

[20] Z. Lu, Y. Wen, and G. Cao, "Information diffusion in mobile social networks: the speed perspective," *Proceedings of IEEE Conference on Computer Communications (IEEE INFOCOM)*, 2014.

[21] Firechat App. http://opengarden.com/about.

[22] Foursquare App. http://foursquare.com/about.

[23] J. Leskovec and A. Krevl, "SNAP Datasets: Stanford large network dataset collection," http://snap.stanford.edu/data, June 2014.

[24] D. Lopez-Pintado, "Diffusion in complex social networks," *Games and Economic Behavior*, vol. 62, no. 2, pp. 573-590, Mar. 2008.

[25] J. Sucec, and I. Marsic, "Hierarchical routing overhead in mobile ad hoc networks, " *IEEE Transactions on Mobile Computing*, vol 3, no. 1, pp. 46-56, Mar. 2004.

[26] N. Al-Nabhan, M. Al-Rodhaan, and A. Al-Dhelaan, "Cooperative approaches to construction and maintenance of networks' virtual backbones for extreme wireless sensor applications," *IEEE Sensors Journal*, vol. 14, no. 11, Nov. 2014.

[27] A. Lancichinetti and S. Fortunato, "Benchmarks for testing community detection algorithms on directed and weighted graphs with overlapping communities," *Physical Review E*, vol. 80, no. 1, July 2009.